\documentclass{article}
\usepackage[utf8]{inputenc}
\usepackage{graphicx}
\usepackage{xcolor}
\usepackage{ulem}
\begin{document}
\begin{center}
{\large \bf{Investigating the Hubble Tension: Effect of Cepheid Calibration}}\\
\vspace{1cm}
Rahul Kumar Thakur$^{1,3}$,
Harish Kumar$^{2}$,
Shashikant Gupta$^{2}$,
Dinkar Verma$^{2}$,
Rahul Nigam$^{1}$
\vspace{1cm}

$^1$ Department of Physics, Birla Institute of Technology $\&$ Science, Pilani  Hyderabad Campus, Hyderabad, 500078, India\\
$^2$ GD Goenka University, Gurugram, 122103, India\\
$^3$ Avantika University, Ujjain, 456006, India\\

\end{center}

\begin{abstract}
Recent observations of Type Ia supernovae (SNe) by SH0ES collaboration (R11 and R16) diverge from the value reported by recent CMBR observations utilising the Planck satellite and application of the $\Lambda CDM$ cosmological model by at least $3 \sigma$. It is among the most challenging problems in contemporary cosmology and is known as the Hubble tension. The SNe Ia in R11 and R16 were calibrated through cepheid variables in three distinct galaxies: Milky Way, LMC, and NGC4258. Carnegie Hubble Program (CHP) observations of type Ia SNe calibrated using the tip of the red giant approach yielded a somewhat different estimate for the Hubble constant. This decreased the Hubble tension from over 3$\sigma$ to below 2$\sigma$. It is a legitimate question to answer whether there are any issues with SNe Ia calibration and to investigate whether the Hubble tension is real or not. We use statistical techniques namely, ANOVA, K-S test, and t-test to examine whether the cepheid calibration is host-dependent. Our analysis shows that (i) both R11 and R16 data suffer from non-Gaussian systematic effects, (ii) $H_0$ values in the sub-samples (different anchor-based) in both R11 and R16 groups are significantly different at a 99\% confidence level, and (iii) neglecting the metal-rich MW sample does not reduce the $H_0$ value significantly, and thus Hubble tension persists. A small reduction in the Hubble constant could be linked to the differences in the host environment. Hence instead of using a single universal relation environment based slope and zero point should be preferred. 

\end{abstract}

\section{Introduction}
The Hubble constant ($H_0$) has been one of the key parameters of modern cosmology since it determines the expansion rate as well as the age of the Universe. Its numerical value also controls the critical density, $\rho_c = 3H_0/8\pi G$, required for the flat geometry of space and many other cosmological parameters such as the physical properties of galaxies and quasars and the growth of large-scale structures.

Measuring $H_0$ with great precision has been among the most important challenges in the last few decades. Various methods have been employed to determine the accurate value of the Hubble constant. Two fundamentally different methods, the Cosmic Microwave Background Radiation (CMBR) \cite{hinshaw2013nine,aghanim2020planck} along with the standard cosmological model and distance ladder-based method using Type Ia supernovae (SNe Ia) provide quite different results, an issue known as the Hubble tension. The observations of Cosmic Microwave Background (CMB) anisotropies provide a global value of  $H_0 = 67.8\pm 0.9 \pm 1.1$ km/S/Mpc. SNe Ia calibrated using the Cepheid variables in different galaxies (SH0ES program), and time delay measurements of gravitationally lensed quasars \cite{wong2020h0licow} provide a higher value which is more than 3$\sigma$ away from that provided by CMBR \cite{riess20113,riess20162,riess2019large,dainotti2021hubble,chen2017determining,thakur2021measurement,thakur2021cosmological,vagnozzi2020new}. Different explanations have been proposed, including a dark component in the early Universe {\cite{slatyer2018early,niedermann2020resolving}} or modifications in the late universe physics \cite{dutta2019cosmology,krishnan2020there}  to resolve the Hubble tension.  

The SH0ES program was designed to refine the SNe Ia calibration using the Hubble Space Telescope (HST) to precisely measure the Hubble constant and determine the equation of the state parameter of dark energy \cite{riess20162}.
HST key project was the first generation attempt to measure $H_0$ by observing Cepheids in various SNe host galaxies. It used the WFPC2 camera, which had a limited range ($\approx 20 $Mpc). SH0ES program, on the other hand, uses the advanced ACS, WFC3, and NICMOS instruments and hence is a “second-generation” attempt to measure $H_0$ using HST. Only suitable SNe Ia, which are spectroscopically normal, have low reddening, and for which observation before maximum brightness is available, have been considered in the SH0ES sample. 

A re-analysis of the Cepheid data that was used to calibrate Type Ia supernovae to determine the Hubble constant $H_0$ was conducted by \cite{perivolaropoulos2021hubble}. 
The empirical Cepheid calibration parameters $R_W$ (Cepheid Wesenheit colour-luminosity parameter) and $M^W_H$ (Cepheid Wesenheit H-band absolute magnitude) were allowed to vary for each unique galaxy rather than a universal value. Additionally, the possibility of two universal values of these parameters arises: one for large galactic distances ($D > D_c$) and one for small galactic distances ($D<D_c$), where $D_c$ is a critical transition distance. Their findings show a three-$\sigma$ disparity between the values of the low and high galactic distance parameters. 

To deduce the value of $H_0$, \cite{mortsell2021hubble} analysed the SH0ES data \cite{riess20162}. Instead of enforcing a universal colour-luminosity relation to correct the Cepheid magnitudes, they employ a data-driven approach in which the optical colours and near-infrared magnitudes of the Cepheids are used to derive individual colour luminosity relations for each Type Ia supernova host and anchor galaxy. The Milky Way anchor is in tension with the NGC 4258 and the Large Magellanic Cloud anchors by a factor of 2.1–3.1 $\sigma$ when individual extinction rules are taken into account and the correlated nature of $H_0$ is taken into consideration.

In the present work, we wish to address the following questions: (i) Are there systematic issues in the $H_0$ measurement of the SH0ES program? (ii) Are there appreciable differences in the $H_0$ values produced from SNe Ia calibration using various anchors?
If yes, what are the implications? The rest of the paper is organized as follows. Section~\ref{sec:data-method} describes the methodology and the data employed. In section~\ref{sec:results}, we present the main results of our analysis. 
This is followed by Section~\ref{sec:disc}, wherein we present a summary and discussion.

\section{Data and Methods}
\label{sec:data-method}
\subsection{Data} 
For our analysis, we have used data from \cite{riess20113} (R11) and \cite{riess20162} (R16). The measurement of $H_0$ in both R11 and R16 relies on (i) the calibration of Cepheid variables in the Milky Way (MW), LMC, and NGC4258, and (ii) SNe Ia light curve calibration through these Cepheid stars. Along with the photometric observations, the geometric distances through parallax methods have also been confirmed for the MW sample of Cepheids (13 in R11 and 15 in R16). The parallax measurements were made using HST and Hipparcos, and are available in \cite{van2007cepheid,benedict2007hubble,riess2014parallax,casertano2016parallax}. One should note that the MW Cepheid calibration provides the highest value of $H_0$ among all three samples. 
Geometric distances through detached eclipsing binaries (DEBs) have been used along with the photometry to calibrate LMC cepheids. DEBs composed of both early type as well as late type stars have been used to measure the distance to the LMC \cite{pietrzynski2013eclipsing}. 
Thus, there are three subsets of data in both R11 and R16. Each subset in R11 has 15 data points which are presented in Table 4 of \cite{riess20113}. The overall 3\% accuracy in the measurement of $H_0$ has been claimed in R11.
R16 is a comparatively more recent and more precise data set. The authors claim a 2.4\% precision in the measurement of $H_0$. Each sample of MW, LMC, and NGC4258 contains 22 data points which are presented in Table 8 of \cite{riess20162}.

\subsection{Methodology}
\label{sec:method}
We have attempted to address two questions in this work. First, we  attempt to answer the issue related to the systematics in R11 and R16 data. Assuming that $H_0^{true}$ represents the true value of the Hubble constant, its measurement can be expressed as 
\begin{equation}
    H_0^{obs} =  H_0^{true} \pm \epsilon \, ,
\end{equation}
where $\epsilon$ is the error in the measurement which could be a combination of various complex processes. If we suppose there are no systematic influences throughout the measuring process, then $\epsilon$ is purely statistical with zero mean. Corresponding to the $i^{th}$ measurement, one can define, \cite{singh2016measurement,thakur2021cosmological} 
\begin{equation}
    \chi_i = \frac{H_{0,i}^{obs} -  H_0^{true}}{ \epsilon_i}
    \label{eq:chi}
\end{equation}
Since uncertainty in each measurement arises from the combination of various processes, $\epsilon_i$ can be treated as a combination of random variables. The central limit theorem suggests that $\epsilon_i$ should follow normal distribution, and hence $\chi_i$ should obey the standard normal distribution. However, if systematic influences were present, it may deviate from that. In this case, the null hypothesis ($H_{Null}$) would be that the data in vector $\chi$ are drawn from the standard normal distribution, against the alternative ($H_A$) that it does not come from such a distribution. 
\begin{equation}
    H_{Null}: \chi \textrm{ follows the standard normal distribution.}
    \label{eq:null-1}
\end{equation}
And, 
\begin{equation}
    H_{A}: \chi \textrm{ does not follow a normal distribution.}
\end{equation}
We employ the Kolmogorov-Smirnov (K-S) test to ensure that the uncertainties in the measurement process obey normal distribution. The K-S test calculates the cumulative distribution (CDF) from the sample and compares it with CDF of standard normal distribution \cite{sheskin2004handbook}. The comparison is based on the distance $D$ mathematically defined as
\begin{equation}
    D = max |F_0(x)-F_r(x)| \,    
\end{equation}
where $F_0(x)$ is the observed cumulative frequency distribution of the random sample of n observations and $F_r(x)$ is the theoretical frequency distribution. Depending on the confidence level and the distance $D$ test returns a parameter $h$. $h=1$ is returned if the null hypothesis is rejected, while $h=0$ means failing to reject the null hypothesis. The K-S test is quite useful as it is sensitive to the difference in both the location as well as the shape of the distribution.

Next, we consider the following question: do different  measurements using the same method, but different anchors yield different $H_0$ values? This is quite important because the same calibration techniques were used for all three samples in R11 as well as in R16 \cite{riess20113,riess20162}. If the difference is statistically significant, what do they imply? They possibly demonstrate that either the cepheid population is not uniform across galaxies or there are systematics that depends on the host galaxy environment. 
To determine if the differences are significant, we employ the analysis of variance (ANOVA) \cite{sheskin2004handbook,gonzalez2012fuzzy}.
One-way ANOVA, a statistical tool is used to compare the variances of three or more samples. The  ANOVA test aims  to check for variability within the data samples as well as the variability among the samples. We have three subgroups belonging to NGC4258, MW, and LMC as anchors in both the data samples, i.e., R11 and R16. In this case, our null hypothesis ($H_{Null}$)  would be that the mean of the populations from which the three samples are drawn, are equal.
\begin{equation}
    H_{Null}:  \mu_1 = \mu_2 = \mu_3 \, ,
    \label{eq:null-2}
\end{equation}
where $\mu_1, \mu_2,$ and $\mu_3$ are the mean of the populations from which $\chi_i$ of NGC4258, MW, and LMC are drawn. The alternative hypothesis ($H_A$) would be that the mean of at least one of the populations is different. It is to be noted that the uncertainties in the measurement of $H_0$ have been incorporated in $\chi_i$, thus the measurement errors are included in the subsequent analysis. The primary assumptions underlying ANOVA are that the measurements are independent and obey a normal distribution. Nevertheless, ANOVA is quite robust as it tolerates violation of the normality assumption well. To test the null hypothesis against the alternative, ANOVA uses the F-value, which is the ratio of between group variability and within group variability. The between group variability is the variance of sample means. 
On the other hand, within group variability is the variance by considering all the data samples together. F can be calculated using the following formula 
\begin{equation}
F = \frac{\Sigma_{i=1}^{k} (\overline{X_i} - \overline{\overline{X}} )^2/(k-1)}{\Sigma_{j=1}^{N} \Sigma_{i=1}^{k} ({X_j} - \overline{X_i} )^2/(N-k)} \, ,
\label{eq:MS}
\end{equation}
where $(\overline{X_i})$ is the mean of $i^{th}$ data sample and $k$ is number of samples. $(\overline{\overline{X}})$ is the grand mean (mean of sample means). N is the total number of observations of all the samples, i.e., $N = \Sigma_{i}^k n_i$, where $n_i$ is the number of observations in $i^{th}$ sample. 
The $F_{cal}$ calculated above is compared with the critical value ($F_{critical}$) obtained from the standard F-distribution tables. If the calculated F value is greater than $F_{critical}$ then we reject the null hypothesis. 

In addition to ANOVA, we also perform two more tests to confirm which anchor galaxy is different from the other two if any:
(i) two sample t-test to compare the means of two groups of $\chi_i$ in a pair of two anchor galaxies, (ii) two sample K-S test to compare the CDF of two groups of $\chi_i$ in a pair of two anchor galaxies. Any sample that differs from the other two samples would not be used for further analysis. From the remaining samples best-fit is again calculated using the maximum likelihood method. The best-fit obtained in this way can be compared with the Planck result to constrain the Hubble tension.

\begin{center}
\begin{table}
\centering
\begin{tabular}{ |c|c|c|c|c|} 
 \hline
S.No.  & Anchor & Best-fit $H_0$ & Standard deviation & No. of data points \\ 
 \hline
 1. & NGC4258 & 74.8
 & 0.91
 & 15\\
 \hline
 2. & MW & 75.66 & 0.95
 & 15\\
 \hline
 3. & LMC & 71.31
 & 1.04
 & 15\\
 \hline
\end{tabular}
\caption{Mean and standard deviation of $H_0$ values (R11) obtained from different anchors, i.e., NGC4258, MW, and LMC.}
\end{table}
\end{center}

\begin{center}
\begin{table}
\centering
\begin{tabular}{ |c|c|c|c|c|} 
 \hline
S.No.  & Anchor & Best-fit $H_0$ & Standard deviation  & No. of data points \\ 
 \hline
 1. & NGC & 72.25 & 0.82 & 22\\
 \hline
 2. & MW & 76.18 & 0.97 & 22\\
 \hline
 3. & LMC & 72.04 & 0.78 & 22\\
 \hline
\end{tabular}
\caption{Mean and standard deviation of $H_0$ values (R11) obtained from different anchors, i.e., NGC4258, MW, and LMC. Best-fit value of $H_0$ for MW clearly stands out.}
\label{Table:MeanH0}
\end{table}
\end{center}

\begin{figure}
\centering

\includegraphics[height=8.0cm]{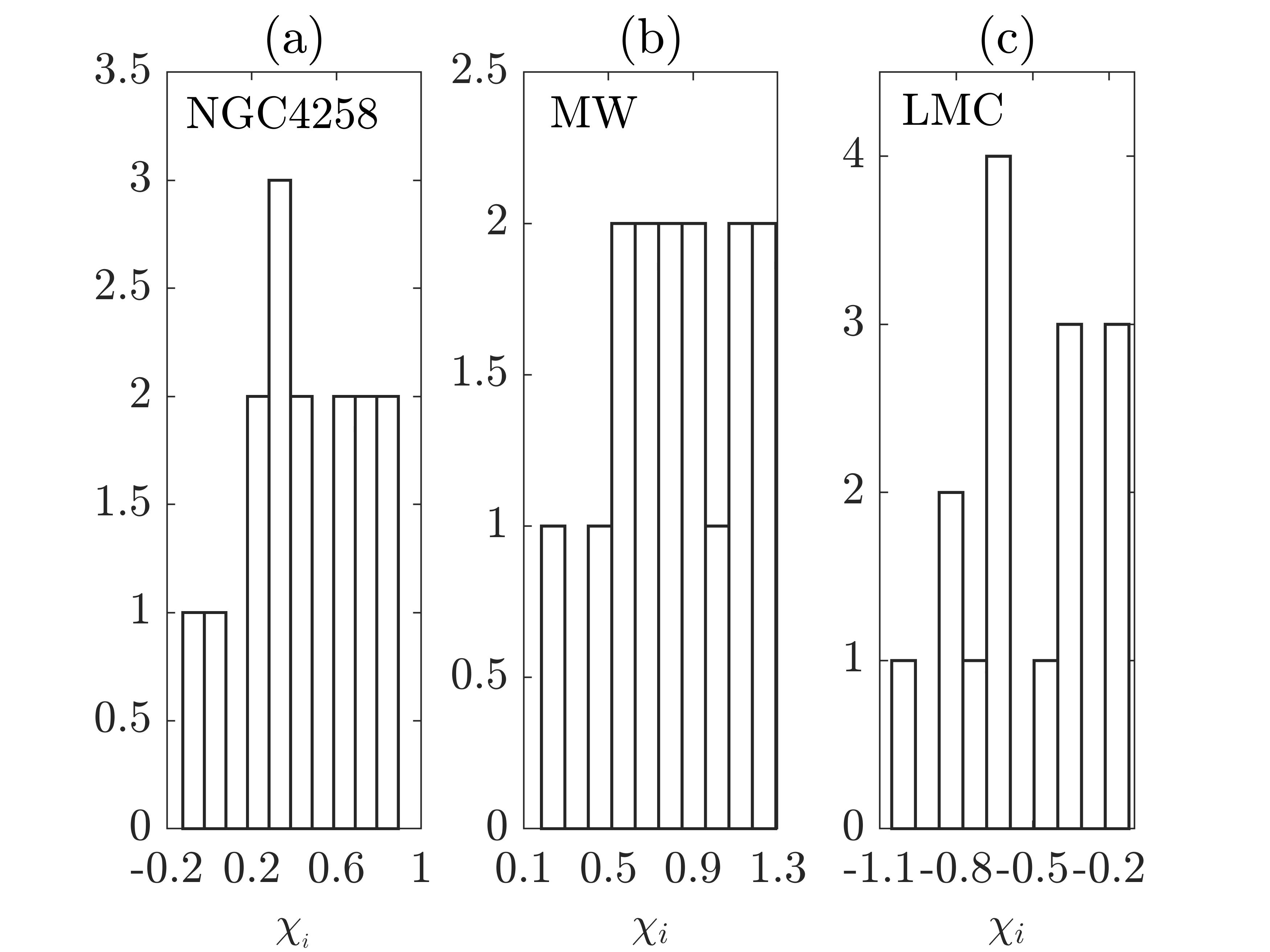}
\caption{Frequency distribution of $\chi\textcolor{blue}{_i}$ values from R11 obtained from different galaxies: (a) NGC4258, (b) MW, and (c) LMC.}
\label{fig:Hist-R11}
\end{figure}

\begin{figure}
\centering

\includegraphics[height=8.0cm]{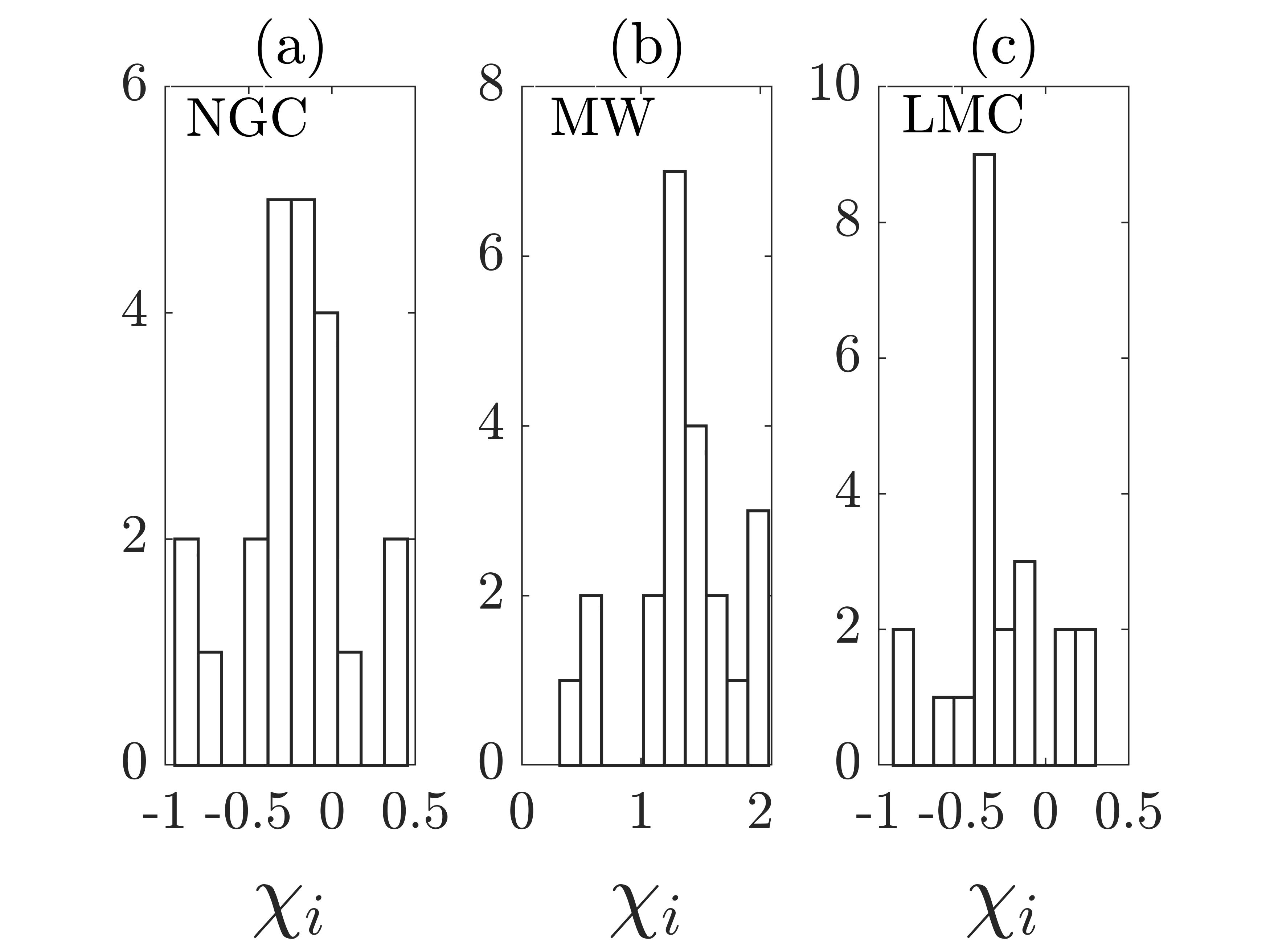}
\caption{Frequency distribution of $\chi\textcolor{blue}{_i}$ values from R16 obtained from different galaxies: (a) NGC4258, (b) MW, and (c) LMC.}
\label{fig:Hist-R16}
\end{figure}


\section{Results}
\label{sec:results} 
We have two data sets, namely the R11 containing three subsets of 15 data points each, and R16 which also has three subsets of 22 data points each. The three subsets in both R11 and R16 belong to the MW, LMC, and NGC 4258 for Cepheid calibration. 
 
We obtain the best-fit value of the Hubble constant for each subset by minimizing $\chi^2$. The best fit values for various subsets of R11 and R16 are presented in tables 1 and 2, respectively. Both tables make it evident that the MW calibration offers a greater value of $H_0$ in comparison to the LMC and NGC4258.

Now we compute $\chi_{\textcolor{blue}{i}}$ from the observed $H_0$ for each subset as defined in Eq.~\ref{eq:chi}. Since $H_0^{true}$ is not available, we use the best-fit value of $H_0$. As discussed before, $\epsilon_i$ is the error in the measurement. Figures 1 and 2 show a histogram of  $\chi_{\textcolor{blue}{i}}$ for each subset. To comprehend the systematics involved in the measurement process, we need to test the null hypothesis defined in Eq.~\ref{eq:null-1}. We first apply the one-sample K-S test for this purpose. The results presented in tables 3 and 4 clearly show that the null hypothesis is rejected in all cases, i.e., measurements do not follow a normal distribution. A comparison of the cumulative distribution of $\chi_{\textcolor{blue}{i}}$ defined in Eq.~\ref{eq:chi} with the standard normal distribution is plotted in figures 3 and 4. The differences between the empirical and the observed distribution are quite clear. 

The next step is to find out whether the various samples differ significantly from one another. In other words, we will test the null hypothesis defined in Eq.~\ref{eq:null-2}. We could have used the Student's t-test if there were only two samples. However, because each data set has three subgroups, we use the ANOVA, whose fundamental structure has been covered in Section~\ref{sec:method}. Similar to the K-S test, ANOVA has also been applied on the $\chi_{i}s$ which carry the information about both the measured value of $H_0$ and the error involved in the measurement. The result of ANOVA for R11 and R16 have been presented in tables 5 and 6, respectively. The $F$ value calculated using Eq.~\ref{eq:MS} is much greater than $F_{crit}$ obtained from the table. Thus the null hypothesis that the three subgroups within the R11 data have the same mean has been rejected at a 99\% confidence level. The p-value is $10^{-15}$, indicating that the chances of above rejection are negligible. Similarly, for R16 as well, $F_{cal}$ is much large compared to the $F_{crit}$ at 99\% confidence level, indicating that the null hypothesis is rejected. $p$ value in this is of the order of $10^{-24}$, which shows that the probability of getting this rejection by chance is again negligible. This implies that the measurements using distinct anchors provide different values and the differences are statistically significant.

The ANOVA test's outcome shows differences in at least one of the sample means. To confirm the outcomes of the ANOVA test and to determine which of the three samples is different from the other two, we further apply the following two tests on various sample pairs: (i) two-sample K-S test and (ii) two-sample paired t-test. Since R16 data is more recent and have more data point, further analysis is done with R16 only. Three pairs of samples are present: MW-LMC, MW-NGC4258, and LMC-NGC4258. The result of the K-S test is presented in Table~\ref{table:kspair} which shows that the MW sample conflicts with the other two samples. The null hypothesis that the two samples in a pair follow the same distribution is rejected ($h=1$) when the MW is one of the samples in the pair. On the other hand, the test fails to reject the null hypothesis for the LMC-NGC4258 pair. A comparison made among CDFs of $\chi_i$ for each sample pair has been shown in Figure~\ref{fig:kspair}. It depicts that the MW sample is separated greatly from the other two samples. Next, we apply a two-sample paired t-test on different sample pairs. The results of the t-test are shown in Table \ref{Table:t-test}, and they also show that the MW sample conflicts with the other two samples. The means of $\chi_i$ for the three anchors have been shown in Figure~\ref{fig:meanXi}. The MW sample mean, $\left< \chi \right>_{MW}$ is quite different from the other two and is around $3\sigma$ away from zero. 
Thus, the results of the K-S test and the t-test are in sync with each other. 
The MW values in the data table of R16 MW are relatively high compared to the other two anchors. One of the reasons for this difference could be that the metallicity of MW is significantly higher than that of LMC and NGC4258. LMC Cepheids are ideal for PLR calibration due to their chemical homogeneity. Also, their metallicity is neither so high nor so low \cite{romaniello2021iron}.

Now that we are certain that MW Cepheids are not ideal for PLR calibration, we discard the MW sample. Now only the LMC and NGC4258 samples are considered together to form a new sample which consists of 44 measurements of $H_0$. We calculate the best-fit value of $H_0$ for this sample by minimising $\chi^{2}$. The probability distribution of $H_0$ is plotted in figure~\ref{fig:prob}. The CCHP and Planck measurements have also been shown for comparison in the same figure. The best-fit value $72.61 \pm 0.38$ is quite different from the other two. This indicates that the Hubble tension is real even if we exclude the MW sample, which yields a greater value of $H_0$. The implications of the above findings are discussed in the next section.
\begin{figure}
\centering
\includegraphics[width=14.0cm]{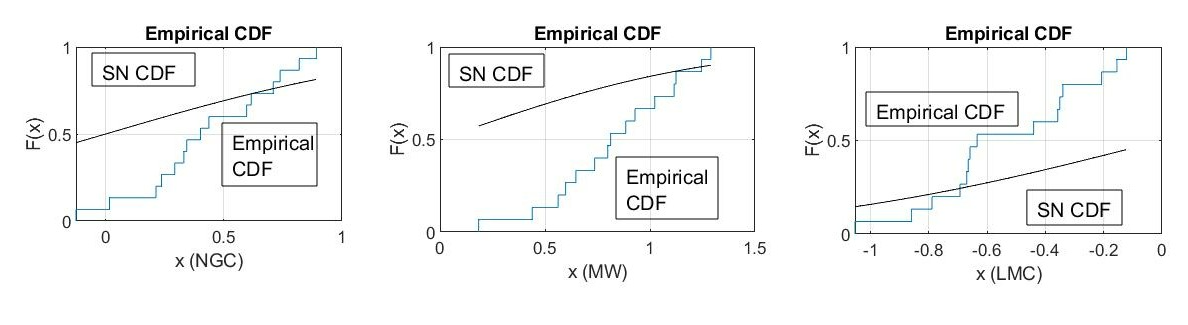}
\caption{A comparison of cumulative distribution of $\chi\textcolor{blue}{_i}$ (R11) with that of standard normal distribution for NGC4258, MW, and LMC. }
\end{figure}
\begin{figure}
\centering
\includegraphics[width=14.0cm]{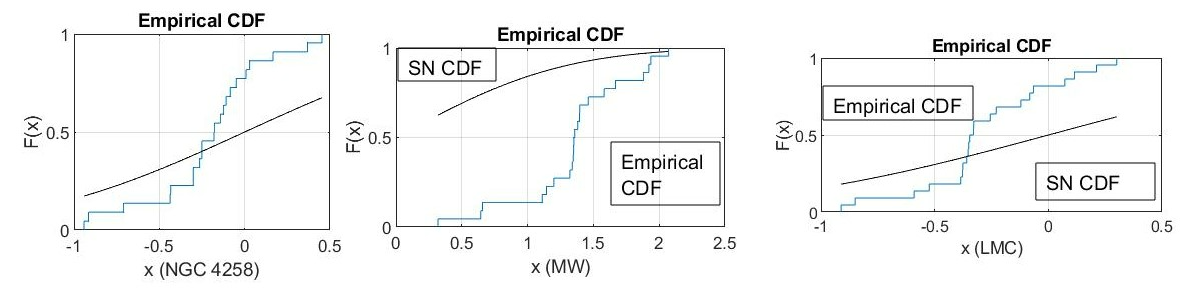}
\caption{A comparison of cumulative distribution of $\chi\textcolor{blue}{_i}$ (R16) with that of standard normal distribution for NGC4258, MW, and LMC. }
\end{figure}

\begin{center}
\begin{table}
\centering
\begin{tabular}{ |c|c|c|c|c|} 
 \hline
S. No.  & Anchor & h value & No. of data points & Result \\ 
 \hline
 1 & NGC4258 & 1 & 15 & R \\
 \hline
 2 & MW & 1 & 15 & R \\
 \hline
 3 & LMC & 1 & 15 & R\\
 \hline
\end{tabular}
\caption{Results of the K-S test on distribution of $\chi_i$ of R11. The null hypothesis is consistently rejected, as indicated by the letter R in the last column.}
\end{table}
\end{center}

\begin{center}
\begin{table}
\centering
\begin{tabular}{ |c|c|c|c|c|} 
 \hline
S. No.  & Anchor & h value & No. of data points & Result \\ 
 \hline
 1 & NGS4258 & 1 & 22 & R \\
 \hline
 2 & MW & 1 & 22 & R \\
 \hline
 3 & LMC & 1 & 22 & R\\
 \hline
\end{tabular}
\caption{Results of one sample K-S test on distribution of $\chi_i$ of R16 to test the normality. The null hypothesis is consistently rejected, as indicated by the letter R in the last column.}
\end{table}
\end{center}
\begin{figure}
\centering
\includegraphics[width=14.0cm]{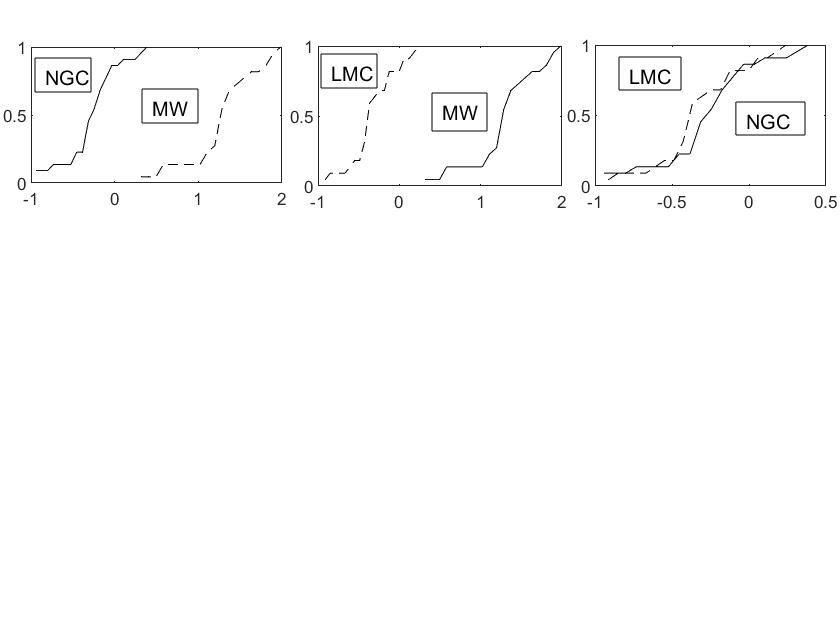}
\vspace{-7cm}
\caption{A comparison of cumulative distribution of $\chi_i$s for various pairs among NGC4258, MW, and LMC (R16).}
\label{fig:kspair}
\end{figure}

\begin{center}
\begin{table}
\centering
\begin{tabular}{ |c|c|c|c|c|c|c|c|c|} 
 \hline
 Variation & SS & f & Mean SS & $F_{cal}$  & $F_{crit}$ & $p$ value  \\
 \hline
 Between group & 14.68
 & 2 & 7.34 & 85.70
 & 3.21 & $1.5*10^{-15}$ \\
 \hline
 Within group & 3.59839 & 42
 & 0.0857 & - & - & -  \\
 \hline
 
\end{tabular}
\caption{Results of one-way ANOVA on $\chi_i$ values for different samples (R11). SS and $f$ stand for sum of squares and Degrees of freedom respectively. $F_{cal}>F_{crit}$ at 99\% confidence level shows rejection of null hypothesis. Very small p-value indicates that the probability of rejection by chance is very small.}
\end{table}
\end{center}

\begin{center}
\begin{table}
\centering
\begin{tabular}{ |c|c|c|c|c|c|c|c|c|} 
 \hline
 Variation & SS & f & Mean SS & $F_{cal}$  & $F_{crit}$ & $p$ value  \\
 \hline
 Between group & 37.65
 & 2 & 18.82 & 144.49
 & 4.95 & $2.91 \times 10^{-24}$\\
 \hline
 Within group & 8.21 & 63
 & 0.130 & - & - & -  \\
 
 \hline
\end{tabular}
\caption{Results of one-way ANOVA on $\chi_i$ values for different samples (R16). $F_{cal}>F_{crit}$ at 99\% confidence level shows rejection of null hypothesis. Very small p-value indicates that the probability of rejection by chance is negligible.}
\label{Table:t-test}
\end{table}
\end{center}
\begin{center}
\begin{table}
\centering
\begin{tabular}{ |c|c|c|c|c|} 
 \hline
S. No.  & Anchor & h value & No. of data points & Result \\ 
 \hline
 1 & NGC4258-MW & 1 & 22 & R \\
 \hline
 2 & MW-LMC & 1 & 22 & R \\
 \hline
 3 & LMC-NGC4258 & 0 & 22 & NR\\
 \hline
\end{tabular}
\caption{Results of paired t-test on the $\chi_i$'s for different pairs of anchor galaxies (R16). The test returns h=1 when one of the samples is from the MW, indicating the null hypothesis is rejected. Clearly, MW sample stands out.}
\label{tabletspair}
\end{table}
\end{center}
\begin{center}
\begin{table}
\centering
\begin{tabular}{ |c|c|c|c|c|} 
 \hline
S. No.  & Anchor & h value & No. of data points & Result \\ 
 \hline
 1 & NGC4258-MW & 1 & 22 & R \\
 \hline
 2 & MW-LMC & 1 & 22 & R \\
 \hline
 3 & LMC-NGC4258 & 0 & 22 & NR\\
 \hline
\end{tabular}
\caption{Results of two-sample K-S test on the $\chi_i$s for different pairs of anchor galaxies (R16). The test returns h=1 when one of the samples is from MW, indicating the null hypothesis is rejected. The results are in agreement with the paired t-test and the MW sample stands out.}
\label{table:kspair}
\end{table}
\end{center}

\begin{center}
\begin{table}
\centering
\begin{tabular}{|c|c|c|c|} 
 \hline
 Data Set  & Best-fit of $H_0$ & Standard deviation & No. of the data points  \\
 \hline
 NGC \& LMC  & 72.61 km/s/Mpc  & 0.38 & 44 \\
 \hline
\end{tabular}
\caption{ Best-fit value of $H_{0}$ using the combination of NGC4258 and LMC samples of R16.}
\label{Table:margin}
\end{table}
\end{center}
\begin{figure}
\centering
\includegraphics[width=12.0cm]{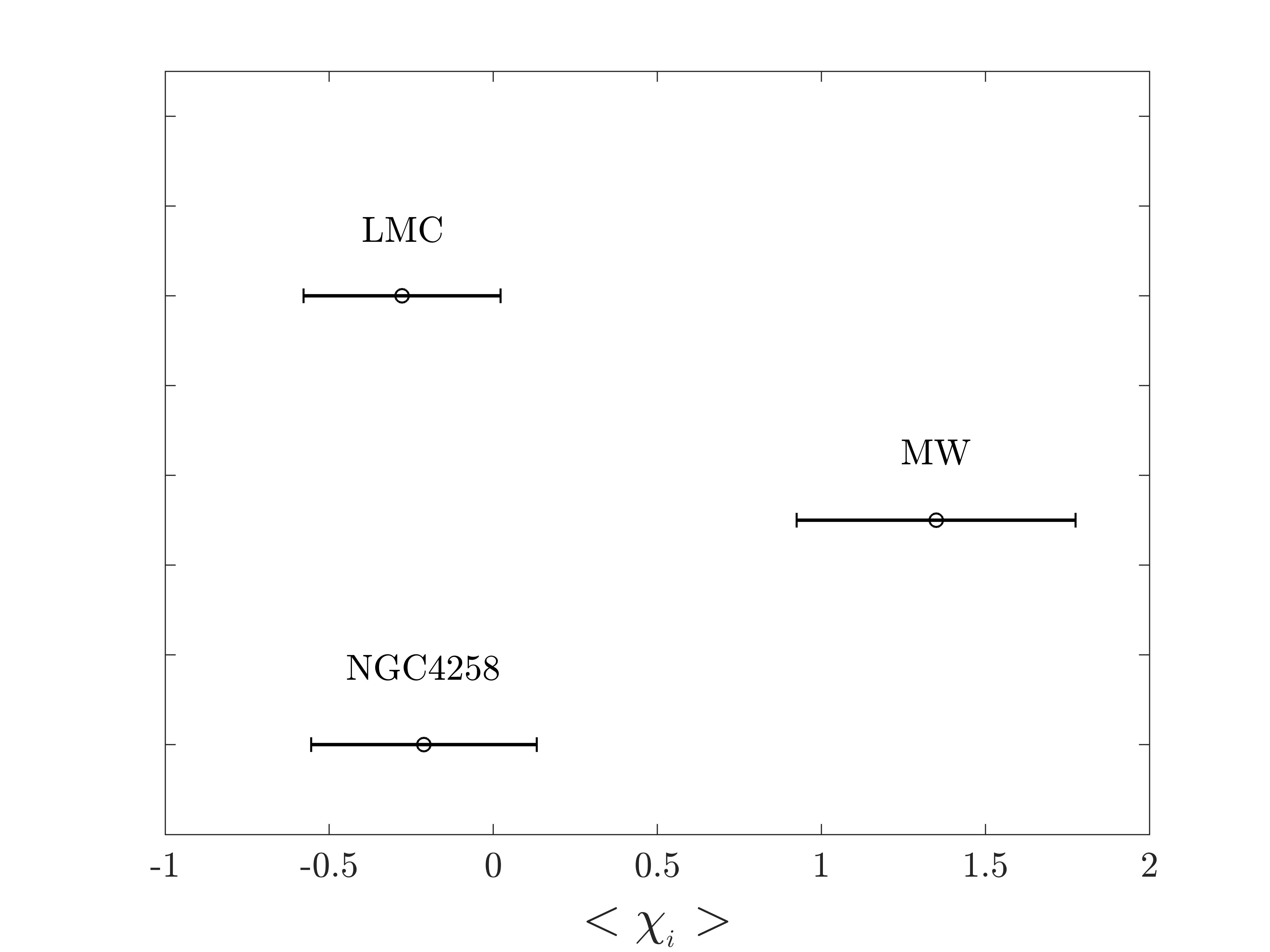}
\caption{The mean of $\chi_i$ for different anchors (R16). Horizontal line shows the standard deviation. $\chi_i$ mean for NGC4258 and LMC are close to each other. MW mean is quite different from the other two anchors and it is around $3\sigma$ away from zero.}
\label{fig:meanXi}
\end{figure}
\begin{figure}
\centering
\includegraphics[width=10.0cm]{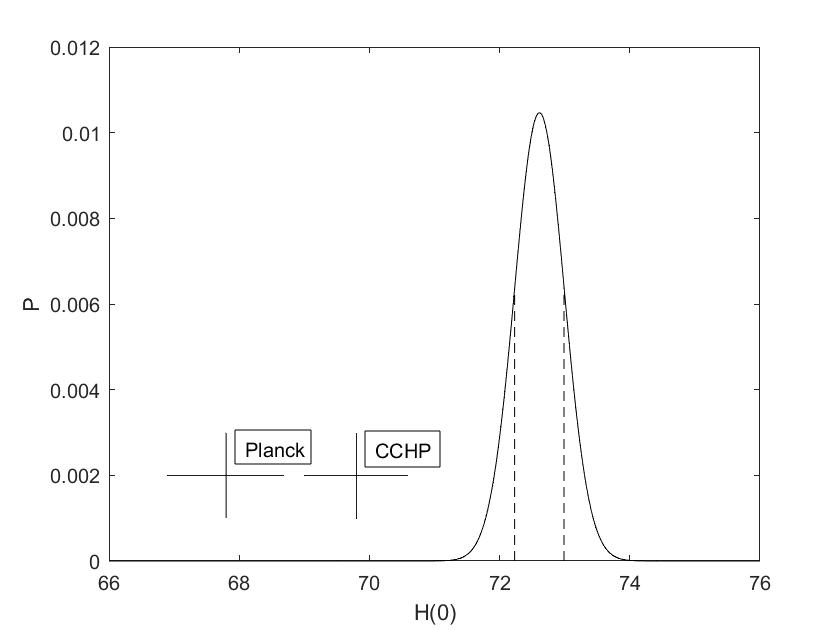}
\caption{Distribution of posterior probability of $H_0$ for the combination of NGC and LMC data. Vertical lines (dash) show the $1\sigma$ level. Both the Planck and CCHP results are outside of the probability distribution.}
\label{fig:prob}
\end{figure}
\section{Summary and Discussion}
\label{sec:disc}
Hubble constant measurement with great precision has been a challenging issue for the last few decades. We have presented a statistical analysis of the Hubble constant measurements using SNe Ia. The data has been taken from \cite{riess20113,riess20162}, one of the key data sets responsible for the Hubble tension. The SNe Ia in R11 and R16 have been calibrated using the cepheid variables in three anchor galaxies: MW, LMC, and NGC4258. The first inference we draw from this work is that both the data sets (R11 and R16) suffer from systematic effects. Another critical inference is about the differences among the different subgroups within the data. The value of the Hubble constant provided by distinct anchors is significantly different. We propose that the difference could be due to the systematic uncertainties arising from the calibration of the P-L relation of Cepheids.

The differences mentioned above are attributed to various systematic issues such as Cepheid mass discrepancy and substantial impact on a star’s evolutionary course due to surface rotation \cite{maeder2000evolution,ekstrom2012grids,georgy2013populations}. Moreover, both theoretical \cite{neilson2015occurrence,mor2017constraining} and observational \cite{szabados2003cepheids,kervella2019multiplicity} studies
show that more than 60\% of the Cepheids are in binary systems. Cepheids with unresolved binary counterparts appear brighter since radiation from the companion also contributes to the overall brightness of the binary system. The observations in the various wavelength bands will be affected differently depending on the companion’s evolutionary stage 
Binary components might have different effective temperatures, which impacts how reddening values are estimated. Therefore, Cepheids with an unresolved companion can change the PLR’s slope and zero point. Furthermore, HST observations indicate that around 44\% of all binary Cepheids are triple systems \cite{evans2005high}; their detection is essential for the mass determination of binary components, as the inclusion of a third component results in erroneous estimates.

The calibrations based on Cepheids P-L relation in different galaxies provide significantly different $H_0$ values. Hence the populations of Cepheid stars may be different in the anchor galaxies due to the metallicity, age, and other environmental factors. This is also supported
by the observations of \cite{perivolaropoulos2021hubble,mortsell2021hubble,gieren2018effect,romaniello2021iron,romaniello2008influence}. It should also be noted that the effect of metallicity on P-L relation is sensitive to the observation band. It has a significant effect in the optical band while almost no effect in IR. Theoretical studies predict fainter optical magnitudes for metal-rich cepheids compared to metal-poor ones. On the other hand, observations show that metal-rich MW cepheids are brighter than metal-poor LMC Cepheids. As suggested by \cite{perivolaropoulos2021hubble,mortsell2021hubble}, instead of using a universal relation, different slopes and zero points in different galaxies may improve the accuracy of $H_0$ measurements. We found that if we neglect MW data and consider only LMC and NGC data,
$H_0$ values are slightly smaller. However, it does not match Planck’s result and thus indicates that Hubble tension is real and needs to be resolved.

\section*{Acknowledgments}
SG thanks SERB for financial assistance (EMR/2017/003714).
\vspace{-1em}

\end{document}